\documentstyle[12pt]{article}
\begin{document}
\tolerance=5000
\def\be{\begin{equation}}
\def\ee{\end{equation}}
\def\bea{\begin{eqnarray}}
\def\eea{\end{eqnarray}}
\def\nn{\nonumber \\}
\def\cF{{\cal F}}
\def\det{{\rm det\,}}
\def\Tr{{\rm Tr\,}}
\def\e{{\rm e}}
\def\etal{{\it et al.}}
\def\erp2{{\rm e}^{2\rho}}
\def\erm2{{\rm e}^{-2\rho}}
\def\er4{{\rm e}^{4\rho}}
\def\etal{{\it et al.}}

\  \hfill
\begin{minipage}{3.5cm}
OCHA-PP-150 \\
NDA-FP-72 \\
\end{minipage}

\vfill

\begin{center}
{\large\bf Dilatonic Brane-World Black Holes,
 Gravity Localization and Newton Constant}

\vfill

{\sc Shin'ichi NOJIRI}\footnote{nojiri@cc.nda.ac.jp},
{\sc Octavio OBREGON}$^{\clubsuit}$\footnote{octavio@ifug3.ugto.mx}, \\
{\sc Sergei D. ODINTSOV}$^{\spadesuit}$\footnote{
odintsov@mail.tomsknet.ru}, 
{\sc Sachiko OGUSHI}$^{\heartsuit}$\footnote{
JSPS Research Fellow,
g9970503@edu.cc.ocha.ac.jp}
\\

\vfill

{\sl Department of Mathematics and Physics \\
National Defence Academy,
Hashirimizu Yokosuka 239, JAPAN}

\vfill
{\sl $\clubsuit$ Instituto de Fisica de la Universidad 
de Guanajuato, Apdo.Postal E-143, 37150 Leon Gto., MEXICO}

\vfill

{\sl $\spadesuit$
Tomsk Pedagogical University, 634041 Tomsk, RUSSIA}

\vfill

{\sl $\heartsuit$ Department of Physics,
Ochanomizu University \\
Otsuka, Bunkyou-ku Tokyo 112, JAPAN}

\vfill

{\bf ABSTRACT}

\end{center}
The family of brane-world solutions of d+1-dimensional dilatonic 
gravity is presented. It includes flat brane with small cosmological 
constant and (anti) de Sitter brane, dilatonic brane-world black 
holes (Schwarzschild-(anti-) de Sitter, Kerr, etc). Gravitational and 
dilatonic 
perturbations around such branes are found. It is shown that near 
dilatonic brane-world black hole the gravity may be localized 
in a standard form. The brane corrections to Newton law are estimated.
The proposal to take into account the dilaton coupled brane matter 
quantum effects is made. The corresponding effective action 
changes the structure of 4d de Sitter wall. RG flow of four-dimensional 
Newton constant in IR and UV is briefly discussed.

\newpage

\section{Introduction}
According to idea clearly expressed 
in ref.\cite{RS} our observed world
could be a brane embedded in
a higher-dimensional space.
Moreover, in the scenario \cite{RS}
 four-dimensional gravity with acceptable
properties may be recovered.
 This fact initiated enormous activity in
the study of different aspects of brane worlds.

In some versions of brane-world scenarios one
can construct the black holes on the branes
\cite{CHR,EHM,ML}. The investigation of their properties
is extremely interesting because this may help to
understand better the fundamental problems of quantum gravity.
For example, one can try to describe from brane point of view
such phenomena as Hawking radiation, Black Hole entropy origin, etc.
It is most likely that brane-world scenario should be realized
within the context of AdS/CFT correspondence (say, in its simplest
form as 5d gauged SG/4d CFT). If it is so one should start
from the scalar-tensor gravity as a bulk theory.
Then the role of scalars (dilaton if only single scalar presents)
should be carefully addressed. In the present paper working in this
direction we construct the family of dilatonic brane-world black holes
(including regular cases ,like de Sitter space) and carefully investigate
their properties, the problem of localization of 4d gravity in such
spaces and
brane corrections to Newton constant.

In the next section we start from d+1-dimensional dilatonic gravity
 with d-dimensional brane vacuum energy and formulate
the corresponding equations of motion. A flat brane solution with
very small cosmological constant is possible. A family of
brane-world black holes solutions (curved branes) is presented.
They correspond to (anti) de Sitter space, Nariai space,
Kerr, and Schwarzschild-de Sitter black holes,etc.
Some their properties are investigated.
In section 3 we look to gravity perturbations
around such backgrounds. Not only graviton but also
dilaton perturbations are found. It is explicitly shown that in some cases
4d gravity may be localized in the same fashion as in ref.\cite{RS}.
Section four is devoted to the study of corrections to Newton constant.
The Newton potential is calculated and it is shown that corrections to
Newton law near branes are very small. In section 5 we analyse the regular
solution
where de Sitter Universe is realised on the brane. Brane vacuum energy is
found. Simple analysis indicates that there may be problems with gravity
localization. The role of quantum effects of brane matter
is investigated in section 6. We suggest to add the conformal anomaly
induced effective action of boundary, dilaton coupled matter to the
complete action. In such a way, Randall-Sundrum compactification
may be naturally fitted with AdS/CFT correspondence.It is shown that with
account of such effective action
there still exists 4d de Sitter wall (our Universe) living in 5d dilatonic
spacetime which is asymptotically AdS. Quantum corrections
change the brane vacuum energies, they become explicitly time-dependent.
RG flow of Newton constant in IR and UV is briefly discussed.
Some resume is given in last section.

\section{Dilatonic black hole solutions in the brane world\label{BHsec}}

In analogy with Randall-Sundrum model \cite{RS},
we start with the following action of the gravity
coupled with dilaton $\phi$:
\bea
\label{S}
S&=& {1 \over 16\pi G_{d+1}}\left[ \int_M d^{d+1}x
 \sqrt{-g}\left( R
 - {1 \over 2}\partial_\mu\phi \partial^\mu\phi
 - V(\phi) \right) \right. \nn
&& \left. - \sum_{i={\rm hid},{\rm vis}}
\int_{B_i} d^d x \sqrt{-\gamma} U_i(\phi)\right]\ .
\eea
Here $M$ is the bulk manifold which usually corresponds to AdS
and $B_{\rm hid}$ and $B_{\rm vis}$ are branes corresponding
to hidden and visible sectors respectively.
$\gamma$ is the metric on the brane induced by the metric $g$
in the bulk.
Here $U_i(\phi)$ corresponds to the vacuum energies
on the branes in \cite{RS}.   One assumes $U(\phi)$ is
dilaton dependent and its form is explicitly given later from
the consistency of the equations of motion.
Some important examples of the dilaton potential are presented in
\cite{CLP}, where $V(\phi)$ is given in terms of the superpotential
$W(\phi)$ :
\be
\label{Vi}
V=\left({\partial W \over \partial \phi}\right)^2
 - {D-1 \over 2(D-2) } W^2 \ ,
\ee
and $W$ has the following form:
\be
\label{Vii}
W=\sqrt{{\cal N} \over 2}g\left({1 \over a_1}\e^{a_1 \phi \over 2}
\pm {1 \over a_2}\e^{a_2\phi \over 2}\right)\ .
\ee
Here $D=d+1$, ${\cal N}$ is the number of the supercharges and
the parameter $g$, $a_1$ and $a_2$ depend on the model features but
$a_1>a_2>0$ in general. As in \cite{CLP}, we only consider the
case of $-$ sign of $\pm$ in (\ref{Vii}). Note that potentials of
above type appear as a result of sphere reduction in M-theory
or string theory \cite{MSB}.

We now assume the metric has the following form:
\be
\label{Mi}
ds^2=dz^2 + \e^{2A(z)}\eta_{ij}dx^i dx^j\ ,
\ee
and $\phi$ only depends on $z$. We also suppose the hidden and
visible branes sit on $z=z_{\rm hid}$ and $z=z_{\rm vis}$,
respectively.
Then the equations of motion are given by
\bea
\label{Ei}
&& \phi''+ (D-1)A'\phi' = {\partial V \over \partial \phi}
+ \sum_{i={\rm hid},{\rm vis}}
{\partial U_i(\phi) \over \partial \phi} \delta(z-z_i)\ , \\
\label{Eii}
&& (D-1)A''+ (D-1)(A')^2 + {1 \over 2}(\phi')^2 \nn
&& \quad = - {V \over D-2}V
 - {D-1 \over 2(D-2)}\sum_{i={\rm hid},{\rm vis}}
U_i(\phi)\delta(z-z_i) \ , \\
\label{Eiii}
&& A'' + (D-1) (A')^2 = - {1 \over D-2} V
 - {1 \over 2(D-2)}\sum_{i={\rm hid},{\rm vis}}
U_i(\phi)\delta(z-z_i) \ .
\eea
Here $'\equiv {d \over dz}$.
For purely bulk sector ($z_{\rm hid}<z<z_{\rm vis}$, as
 $z_{\rm hid}<z_{\rm vis}$), the explicit solutions
are given in \cite{CLP}. Eqs. (\ref{Ei}-\ref{Eiii}) have the
following first integrals (in the bulk sector):
\bea
\label{Iii}
\phi'=\sqrt{2}{\partial W \over \partial \phi}\ ,
\quad A' = - {1 \over \sqrt{2}(D-2)}W\ .
\eea
Near the branes,
Eqs. (\ref{Ei}-\ref{Eiii}) have the following form :
\be
\label{Eiv}
\phi'' \sim {\partial U_i(\phi)\over \partial\phi}\delta (z-z_i)\ ,
\quad A'' \sim -{U_i(\phi) \over 2(D-2)}\delta (z-z_i)\ ,
\ee
or
\be
\label{Eivb}
2\phi' \sim {\partial U_i(\phi)\over \partial\phi} ,
\quad 2A' \sim -{U_i(\phi) \over 2(D-2)}\ ,
\ee
at $z=z_i$.
Comparing (\ref{Eivb}) with (\ref{Iii}), we find
\be
\label{Ev}
U_{\rm hid}(\phi)= 2\sqrt{2}W(\phi)\ ,\quad
U_{\rm vis}(\phi)=- 2\sqrt{2}W(\phi)\ .
\ee
For simplicity, let $z_{\rm hid}=0$, where $\phi=0$
in the solution in \cite{CLP} and we only consider this solution
in the following. Then at $z=z_{\rm vis}$, $\phi$ is negative.
Since the superpotential is given by the exponential of
$\phi$ with positive sign, the vacuum energy $U_{\rm vis}$,
which can be identified with the cosmological constant, can
be small even if $|\phi|$ is not so large.
The result might explain why the cosmological constant is small.
We should note that the visible brane corresponds to the 4d
universe where we live. Then the visible brane part
in the action (\ref{S}) mainly describes the dynamics of the
universe and $U_{\rm vis}$ corresponds to the vacuum energy
or cosmological constant in our universe.

As an extension, one can consider the case that the brane is
curved. Instead of (\ref{Mi}), we take the following metric:
\be
\label{Mi2}
ds^2=dz^2 + \e^{2A(z)}\tilde g_{ij}dx^i dx^j\ ,
\ee
Here $\tilde g_{ij}$ is the metric of the Einstein manifold,
which is defined by
\be
\label{Ein}
\tilde R_{ij}=k \tilde g_{ij}\ ,
\ee
where $\tilde R_{ij}$ is the Ricci tensor given by
$\tilde g_{ij}$ and $k$ is a constant.
Then Eqs.(\ref{Ei}) and (\ref{Eii}) do no change but
one obtains the following equation instead of (\ref{Eiii}):
\be
\label{Eiiib}
A'' + (D-1) (A')^2 = k\e^{2A} - {1 \over D-2} V
 - {1 \over 2(D-2)}\sum_{i={\rm hid},{\rm vis}}
U_i(\phi)\delta(z-z_i) \ .
\ee
Especialy when $k=0$, we get the previous solution for $\phi$,
$A$ and $U_i$. We should note, however, that $k=0$ does not
always mean the brane is flat.
As well-known, the Einstein equations are given by,
\be
\label{A1}
R_{\mu\nu}-{1 \over 2}g_{\mu\nu}R+{1 \over 2}\Lambda g_{\mu\nu}
= T^{\rm matter}_{\mu\nu}\ .
\ee
Here $T^{\rm matter}_{\mu\nu}$ is the energy-momentum tensor of
the matter fields. If we consider the vacuum solution where
$T^{\rm matter}_{\mu\nu}=0$, Eq.(\ref{A1}) can be rewritten
when $d=4$ as (see also \cite{NOcurv})
\be
\label{A2}
R_{\mu\nu}={\Lambda \over 2}g_{\mu\nu}\ .
\ee
If we put $\Lambda=2k$, Eq.(\ref{A2}) is nothing but the equation
for the Einstein manifold. The Einstein manifolds
are not always homogeneous manifolds like flat Minkowski,
(anti-)de Sitter space
\be
\label{AdS}
ds_4^2= -V(r)dt^2 + V^{-1}(r)dr^2+r^2 d\Omega^2,
\qquad
V(r) = 1 - \frac{\Lambda}{6} r^2,
\ee
or Nariai space
\be
\label{Nsol}
ds_4^2={1 \over \Lambda}\left( \sin^2 \chi d\psi^2
- d\chi^2 - d\Omega^2\right)\ .
\ee
but they can be some
black hole solutions like Schwarzschild-(anti-)de Sitter
black hole
\be
\label{SAdS}
ds_4^2= -V(r )dt^2 + V^{-1}(r )dr^2+r^2 d\Omega^2,
\qquad
V(r) = 1- {\tilde G_4 M \over r} - \frac{\Lambda}{6} r^2\ .
\ee
As a special case, one can also consider $k=0$ solution like
Schwarzschild black hole,
\be
\label{schw}
ds_4^2\equiv \tilde g_{ij}dx^i dx^j
=-\left(1 - {\tilde G_4 M \over r}\right)dt^2
+{dr^2 \over \left(1 - {\tilde G_4 M \over r}\right)}
+ r^2 d\Omega^2\ ,
\ee
or Kerr one
\bea
\label{Kerr}
ds_4^2&=& \Delta \tilde A dt^2 - {\Sigma^2 \over \Delta}dr^2
- \Sigma^2 d\theta^2 - {\sin^2\theta \over \tilde A}\left(
d\varphi - \Omega dt\right) \nn
&& \tilde A \equiv {\Sigma^2\left(\Delta - a^2 \sin^2\theta\right)
\over \Sigma^4\Delta - 4a^2 \tilde G_4^2 M^2 r^2 \sin^2 \theta} \ ,
\quad \Omega\equiv{2a \tilde G_4 M r \tilde A \over \Sigma^2} \nn
&& \Delta\equiv r^2 - 2\tilde G_4 M r + a^2\ ,
\quad \Sigma^2 \equiv r^2 + a^2 \cos^2\theta\ .
\eea
In (\ref{SAdS}), (\ref{schw}) and (\ref{Kerr}),
$M$ is the mass of the
black hole on the brane and the effective gravitational
constant $\tilde G_4$ on the 3-brane (here $d=4$)
is given by
\be
\label{schw1}
{1 \over \tilde G_4}
={1 \over G_5}\int_{z_{\rm hid}}^{z_{\rm vis}} dz \e^{(d-2)A}\ .
\ee
In (\ref{Kerr}), the parameter $a$ is related with the angular
momentum $J$ of the black hole on the brane by
\be
\label{AngM}
J = M a\ .
\ee
In these solutions, the curvature singularity
at $r=0$ has a form of line penetrating the bulk 5d universe and
the horizon makes a tube surrounding the singularity.
The singularity and the horizon connect the hidden and visible
branes. Thus we presented a big family of dilatonic solutions
in the brane world.

It would be interesting to see how the black hole looks like in
the bulk (see corresponding discussion in \cite{GKR}).
We now consider the case $|z_{\rm vis}|$ is large but
$z_{\rm vis}<0$ or $z_{\rm vis}>0$. When $z_{\rm vis}$ is
negative and its absolute value $|z_{\rm vis}|$ is large,
the asymptotic behaviour of $A$ is given by
\be
\label{Mp1b}
A \sim \gamma z\ ,\quad \gamma \equiv
{g \over a_1 d \sqrt{{\cal N}}} \ .
\ee
Then since the metric has the form in (\ref{Mi2}),
the typical proper size $l^-_{\rm BH}$ of the black hole
(or the typical proper distance to the horizon) transverse
to the brane would be given by
\be
\label{GKR1}
l^-_{\rm BH}={1 \over \gamma}\ln \left(\gamma r_0\right)
={1 \over \gamma}\ln \left(\gamma \tilde G_4 M\right)\ .
\ee
Then the transverse size $l^-_{\rm BH}$
grows like $\ln M$ when the black hole
mass $M$ increases although the (horizon) size $r_0$ along the
brane is linear to $M$.
This might tell that matter can pass  around the black hole
through the fifth dimension similar to the effect suggested in \cite{GKR}.

On the other hand, we can consider the case that
$z_{\rm vis}$ is positive and large.
The asymptotic behaviour of $A$ when $z$ is large,
is given by \cite{CLP}
\be
\label{PMv}
\e^{2A}\sim \alpha^2 z^{2\beta}\ ,\quad
\alpha\equiv \left\{-c\sqrt{{\cal N} \over 8} g \left({1 \over 4}
a_2\sqrt{{\cal N}} g\right)^{a_1 \over a_2}\right\}^{1 \over d}\ ,
\quad \beta \equiv {a_1 \over a_2 d}\ .
\ee
Then if one defines a new coordinate $w$ by
\be
\label{GKR2}
y={z^{1-\beta} \over \alpha (1-\beta)}\ ,
\ee
the metric in (\ref{Mi2}) can be rewritten in the following form:
\be
\label{Mi2b}
ds^2\sim \alpha^2 \left\{\alpha(1-\beta) y
\right\}^{2\beta \over 1-\beta}\left(
dw^2 + \tilde g_{ij}dx^i dx^j\right)\ .
\ee
Then we can expect the horizon in the bulk is typically given
by $y\sim r_0$ and the typical proper size $l^+_{\rm BH}$ of
the black hole transverse
to the brane would be given by
\be
\label{GKR3}
l^+_{\rm BH}=\left\{\alpha(1-\beta) r_0
\right\}^{1 \over 1-\beta}
=\left\{\alpha(1-\beta) \tilde G_4 M
\right\}^{1 \over 1-\beta}\ .
\ee
Since $a_1=2\sqrt{5 \over 3}$ and $a_2={4 \over \sqrt{15}}$
for $d=D-1=4$ and ${\cal N}=1$ model in \cite{CLP},
we have ${1 \over 1-\beta}={5 \over 3}$ and
the transverse size $l^+_{\rm BH}$
grows like $M^{5 \over 3}$ when the black hole
mass $M$ increases. This would tell that matter cannot pass
through a black hole, which is different from the case when
$z_{\rm vis}\rightarrow - \infty$. Of course, this is a preliminary
qualitative discussion of brane-world black holes properties.

\section{Gravity perturbations}

Our next problem will be the description of the gravity on the brane.
For that purpose one can consider the perturbation from the above obtained
solution by assuming the metric in the following form:
\be
\label{PMi}
ds^2= dz^2 + \e^{2A(z)}\tilde g_{ij}dx^i dx^j+h_{ij}dx^i dx^j\ ,
\quad |h_{ij}|\ll 1\ .
\ee
Here we choose the gauge where $g_{zz}=1$, $g_{zi}=g_{iz}=0$,
$D^i h_{ij}=0$ and $h^{i}_{\ i}=0$.
We should note that the metric $\tilde g_{ij}$ on the brane
is not necessary to be flat but can be any Einstein manifold
(\ref{Ein}). We put $k=0$ for simplicity. Then $A(z)$
is given by solving (\ref{Iii}) \cite{CLP}. From the Einstein
equation, we obtain the following linearized equation:
\be
\label{PMii}
0=\e^{-2A}\tilde \Box^{(0)} h_{ij} + \partial_z^2 h_{ij}
+ (d-4)\partial_z A \partial_z h_{ij}
 -4 \left\{ (d - 1) \partial_z A^2
 + \partial_z^2 A\right\}h_{ij}\ .
\ee
Here $\tilde \Box^{(0)}$ is the d'Alembertian on the brane given
by $\tilde g_{ij}$.
Changing the coordinate $z$ to $\zeta$ by
\be
\label{PMiii}
d\zeta =-\e^{-A}dz\ ,
\ee
one rewrites (\ref{PMii}) in the following form:
\be
\label{PMiv}
0=\tilde \Box^{(0)} h_{ij} + \partial_\zeta^2 h_{ij}
+ (d-5)\partial_\zeta A \partial_\zeta h_{ij}
 -4 \left\{ (d - 2) \partial_\zeta A^2
 + \partial_\zeta^2 A\right\}h_{ij}\ .
\ee
In order to solve Eq.(\ref{PMiv}), we assume the following
form for $h_{ij}$:
\be
\label{PMix}
h_{ij}=\psi(\zeta)\hat h_{ij}(x)
\ee
and assume $\hat h_{ij}(x)$ satisfies
\be
\label{PMx}
\tilde \Box^{(0)}\hat h_{ij}(x)=m^2 \hat h_{ij}(x)\ .
\ee
Here $m$ corresponds to the mass. Then one has
\be
\label{PMivb}
0=m^2\psi + \partial_\zeta^2 \psi
+ (d-5)\partial_\zeta A \partial_\zeta \psi
 - 4\left\{ (d - 2) \partial_\zeta A^2
 + \partial_\zeta^2 A\right\}\psi\ .
\ee

As we have a dilaton field $\phi$, we should consider the
perturbation of $\phi$:
\be
\label{phP1}
\phi = \phi^{(0)} + \varphi\ , \quad |\varphi|\ll
|\phi^{(0)}|\ .
\ee
Here $\phi^{(0)}$ is given by solving (\ref{Iii}).
Linearizing the equation of motion given by the variation
of $\phi$,
\be
\label{phP2}
\Box \phi = {\partial V(\phi) \over \partial \phi}
+ \sum_i {\partial U_i(\phi) \over \partial \phi}
\delta(z-z_i)\ ,
\ee
one gets
\be
\label{phP3}
\e^{-2A}{\tilde\Box}^{(0)}\varphi + \partial_z^2\varphi
+ d\partial_z A \partial_z\varphi
=\left\{{\partial^2 V(\phi) \over \partial \phi^2}
+ \sum_i {\partial^2 U_i(\phi) \over \partial \phi^2}
\delta(x-z_i)\right\}\varphi\ .
\ee
Using the coordinate $\zeta$ defined by (\ref{PMiii})
and assuming $\phi$ in the following form:
\be
\label{phP4}
\varphi(\zeta,x)=\theta(\zeta)\hat\varphi(x)\ ,
\quad {\tilde\Box}^{(0)}\hat\varphi(x)=m^2 \hat\varphi(x)\ ,
\ee
we can rewrite Eq.(\ref{phP3}) as:
\be
\label{phP5}
m^2\theta + \partial_\zeta^2\theta
+ (d-1) \partial_\zeta A \partial_\zeta\theta
=\e^{2A}\left\{{\partial^2 V(\phi) \over \partial \phi^2}
+ \sum_i {\partial^2 U_i(\phi) \over \partial \phi^2}
\delta(z-z_i)\right\}\theta\ .
\ee

In the solutions found in \cite{CLP}, there appear two
branches. In the first branch, $z$ runs from $-\infty$ to $0$,
$\phi$ from $0$ to $+\infty$ and there appears a curvature
singularity at $z=0$. In the second one $z$ runs from $-\infty$
to $+\infty$, $\phi$ from $0$ to $-\infty$ and there does not
appear any curvature singularity. In both of the branches,
the spacetime approaches to AdS.

We now consider the perturbation when $z_{\rm vis}$ is
negative and its absolute value $|z_{\rm vis}|$ is large and the
region $z\leq z_{\rm vis}$. Then we find
the following asymptotic behaviour of $A$ and $\phi$:
\be
\label{Mp1}
A \sim \gamma z\ ,\quad \phi \sim \e^{d\gamma z}\rightarrow 0\ ,
\quad \gamma \equiv {g \over a_1 d \sqrt{{\cal N}}} \ .
\ee
Then $\zeta$ in (\ref{PMiii}) is given by
\be
\label{Mp2}
\zeta={\e^{-\gamma z} \over \gamma}
\ee
and Eq.(\ref{PMivb}) has the following form
\be
\label{PMivc}
0=m^2\psi + \partial_\zeta^2 \psi
+ {5 - d \over \zeta} \partial_\zeta \psi
 - { 4(d - 1) \over \zeta^2 }\psi\ .
\ee
When $m> 0$, the corresponding solution is called Kaluza-Klein
(KK) mode and the solution of $m=0$ corresponds to the graviton
on the brane.
When $m^2=0$, the solution of (\ref{PMivc}) is given by
the power of $\zeta$:
\be
\label{Mp3}
\psi = \zeta^a\ .
\ee
The exponent $a$ can be found by solving the following
algebraic equation:
\be
\label{Mp4}
0=a^2 + (4-d) a - 4(d-1)\ .
\ee
Especially when $d=4$, one gets
\be
\label{Mp5}
a=\pm 2\sqrt{3}\ .
\ee

The existence of the normalizable solution tells
that the gravity is localized near the brane.
This situation does not change even if the brane is flat or
a (4d) black hole spacetime.

When $m^2>0$, the solution of (\ref{PMivc}) is given by
Bessel functions $J_\nu$ and $N_\nu$ ($=Y_\nu$):
\bea
\label{Mp6}
&& \psi(\zeta)=\zeta^b\left(c_1 J_\nu(m\zeta) + c_2 N_\nu (m\zeta)
\right) \nn
&& 1-2b=5-d\ ,\quad b^2 - \nu^2=-4(d-1)\ .
\eea
Here $c_1$ and $c_2$ are constants of integration, which should
be determined by the boundary condition at $z=z_{\rm vis}$:
\bea
\label{MP7}
\partial_z\psi(z=z_{\rm vis})
&=&-\e^{-A(\zeta=\zeta_{\rm vis})}
\partial_\zeta\psi(\zeta=\zeta_{\rm vis}) \nn
&=&-{2\sqrt{2} \over d-3}W\left(\phi(z=z_{\rm vis})\right)
\psi(z=z_{\rm vis})\ .
\eea
Here $\zeta_{\rm vis}$ is the value of $\zeta$ corresponding to
$z=z_{\rm vis}$. The boundary condition comes from the
$\delta$-function behaviour of $\partial_z^2 A$
at $z=z_{\rm vis}$.
Especially for $d=4$, we find
\be
\label{MP8}
b=0\ ,\quad \nu=2\sqrt{3}\ .
\ee
If there is a solution with $m^2<0$, the system becomes
unstable. When $m^2<0$, the solution of (\ref{PMivc}) is given by
modified Bessel functions $I_\nu$ and $K_\nu$:
\bea
\label{Mp9}
&& \psi(\zeta)=\zeta^b\left(c_1 I_\nu(\mu\zeta) + c_2 K_\nu (\mu\zeta)
\right) \nn
&& 1-2b=5-d\ ,\quad b^2 - \nu^2=-4(d-1)\ ,\quad \mu^2=-m^2\ .
\eea
Since $I_\mu$ increases exponentially for large $\zeta$,
$c_1$ must vanish. On the other hand, $K_\nu$ behaves as
$K_\mu(m\zeta)\sim\e ^{-\mu\zeta}$ for large $\zeta$.
When $\zeta$ is large, $z\rightarrow -\infty$, $\phi \rightarrow 0$
and $W\rightarrow \sqrt{{\cal N} \over 2}g\left({1 \over a_1}
- {1 \over a_2}\right)<0$. Therefore there is an unstable solution
which satisfies the boundary condition (\ref{MP7}) if one chooses
\be
\label{Mp9b}
\mu\sim {2\sqrt{2} \over d-3}\e^{A(z=z_{\rm vis})} W
\left(\phi(z=z_{\rm vis})\right)
\rightarrow -{2\sqrt{2} \over d-3}\e^{\gamma z_{\rm vis}}
\sqrt{{\cal N} \over 2}g\left({1 \over a_1}
- {1 \over a_2}\right)\ .
\ee
As $\mu$ vanishes in the limit of $z\rightarrow -\infty$, the
brane will be driven to $z\rightarrow -\infty$.

We now consider the perturbation by the dilaton $\phi$.
When $\zeta$ is large, Eq.(\ref{phP5}) can be rewritten as
\be
\label{phP6}
m^2\theta + \partial_\zeta^2\theta
 - {d-1 \over \zeta} \partial_\zeta\theta
 - {V_0'' \over \gamma^2\zeta^2}\theta
=\e^{2A}\sum_i {\partial^2 U_i(\phi) \over \partial \phi^2}
\delta(z-z_i)\theta\ .
\ee
Here
\be
\label{phP7}
V_0''\equiv \left.{\partial^2 V(\phi) \over \partial \phi^2}
\right|_{\phi=0}
= {1 \over g^2}\left(1 - {a_2 \over a_1}\right)>0\ .
\ee
Then one finds $\theta$ is given by the power of $\zeta$ when
$m^2=0$ or Bessel functions when $m^2>0$. From the
$\delta$-function,  we find that $\theta$ should satisfy the
following boundary condition $\zeta=\zeta_{\rm vis}$ :
\be
\label{phP8}
\partial_\zeta \theta = - {\sqrt{2} \over \gamma\zeta_{\rm vis}}
\left.{\partial^2 W \over \partial\phi^2}\right|_{\phi=0}
\theta\ .
\ee
Since
\be
\label{phP9}
\left.{\partial^2 W \over \partial\phi^2}\right|_{\phi=0}
 =\sqrt{{\cal N} \over 2}{g(a_1 - a_2) \over 4}>0\ ,
\ee
the coefficient $- {\sqrt{2} \over \gamma\zeta_{\rm vis}}
\left.{\partial^2 W \over \partial\phi^2}\right|_{\phi=0}$ is
negative. Then the boundary condition is consistent with the
modified Bessel function $K_\nu$ and the condition could be
satisfied by properly choosing $m^2$. This tells that there exists
an unstable mode corresponding to $m^2<0$.


In the second branch in the solution in \cite{CLP},
we can consider the case that $z_{\rm vis}$ is positive and
large. We also consider the region $z\leq z_{\rm vis}$.
Since the asymptotic behaviour of $A$ when $z$ is large
is given by (\ref{PMv}), one gets
\be
\label{PMvi}
-\zeta\sim {z^{1-\beta} \over 1-\beta}
\ee
and Eq.(\ref{PMiv}) has the following form
\be
\label{PMvii}
0=\tilde \Box^{(0)} h_{ij} + \partial_\zeta^2 h_{ij}
+ {(d-5)\tilde\beta \over \zeta} \partial_\zeta h_{ij}
 -4 \left\{ (d - 2) \tilde\beta^2
 - \tilde\beta \right\}h_{ij}\ .
\ee
Here
\be
\label{PMviii}
\tilde\beta\equiv {\beta \over 1-\beta}\ .
\ee
and we obtain
\be
\label{PMxi}
0=m^2\psi + \partial_\zeta^2 \psi
+ {(d-5)\tilde\beta \over \zeta} \partial_\zeta \psi
 -4 \left\{ (d - 2) \tilde\beta^2
 - \tilde\beta \right\}\psi\ .
\ee
When $m=0$, the solution is given by
\be
\label{PMxii}
\psi = (-\zeta)^a\ .
\ee
Here $a$ is the solution of the following algebraic equation:
\be
\label{PMxiii}
0= a^2 + \left\{(d-5)\tilde\beta -1 \right\} a
 -4 \left\{ (d - 2) \tilde\beta^2
 - \tilde\beta \right\}\ .
\ee
For $d=D-1=4$ and ${\cal N}=1$ model in \cite{CLP},
where $a_1=2\sqrt{5 \over 3}$
and $a_2={4 \over \sqrt{15}}$, we have
\be
\label{PMxiv}
a={4 \pm 2\sqrt{39} \over 3}=5.49\cdots ,
\ - 2.83\cdots
\ee
Since we are considering the case that $z_{\rm vis}$ is
positive and $z\leq z_{\rm vis}$, the first positive $a$
would correspond to a normalizable solution.
The existence of the normalizable solution tells that the gravity
near the brane is localized.

When $m> 0$, the solution of (\ref{PMxi}) is given by
Bessel functions $J_\nu$ and $N_\nu$ ($=Y_\nu$):
\bea
\label{PMxix}
&& \psi(\zeta)=z^b\left(c_1 J_\nu(-m\zeta) + c_2 N_\nu (-m\zeta)
\right) \nn
&& 1-2b=(d-5)\tilde\beta\ ,\quad
b^2 - \nu^2=-4\left\{(d-2)\tilde\beta^2 - \tilde\beta\right\}\ .
\eea
Here $c_1$ and $c_2$ are constants of integration, which should
be again determined by the boundary condition
(\ref{MP7}) at $z=z_{\rm vis}$.
The boundary condition comes from the
$\delta$-function behaviour of $\partial_z^2 A$
at $z=z_{\rm vis}$.
When $\zeta$ ($z$) is large, the Bessel functions
$J_\nu(-m\zeta)$ and $N_\nu(-m\zeta)$ behave as
\bea
\label{PMxxi}
J_\nu(-m\zeta)&\sim& \sqrt{-{2 \over \pi m\zeta}}
\cos\left(-m\zeta - {(2\nu +1)\pi \over 4} \right)
\ ,\nn
N_\nu(-m\zeta)&\sim& \sqrt{-{2 \over \pi m\zeta}}
\sin\left(-m\zeta - {(2\nu +1)\pi \over 4} \right)\ .
\eea
For $d=D-1=4$ and ${\cal N}=1$ model in \cite{CLP},
we have
\be
\label{PMxxii}
b={4 \over 3}\ ,\quad \nu^2 = {4\cdot 39 \over 9}
=(4.16\cdots )^2\ .
\ee
The existence of the unstable mode corresponding to modified
Bessel function $K_\mu(\mu\zeta)$ $(\mu^2=-m^2$
depends on the asymptotic behaviour of
$W$ as in $z_{\rm vis}\rightarrow -\infty$ case. From
(\ref{Vii}) ($-$ sign is chosen), we find $W<0$ when
$\phi\rightarrow -\infty$ and $\e^{A\left(\phi(z)\right)}
W\left(\phi(z)\right)$ vansihes when $z\ (\zeta) \rightarrow
+ \infty$. Therefore an unstable solution given by $K_\mu(\mu\zeta)$
exists and the brane moves to $+\infty$, where the vacuum energy of
the brane vanishes since $\mu\rightarrow 0$ when
$z_{\rm vis}\rightarrow \infty$.

We also consider the perturbation by the dilaton $\phi$.
When $\zeta$ is large, Eq.(\ref{phP5}) can be rewritten as
\be
\label{phP6b}
m^2\theta + \partial_\zeta^2\theta
 +{\tilde\beta (d-1) \over \zeta} \partial_\zeta\theta
 + {\tilde V_0 \over \gamma^2\zeta^2}\theta
=\e^{2A}\sum_i {\partial^2 U_i(\phi) \over \partial \phi^2}
\delta(z-z_i)\theta\ .
\ee
Here
\be
\label{phP7b}
\tilde V_0\equiv  {1 \over 2} g^2
\left(1 - {a_1 \over d a_2}\right)^{-2}\left({a_2 \over a_1}\right)
\left({1 \over 4}a_2\sqrt{{\cal N}} g\right)^{-2}\ .
\ee
Then we find $\theta$ is given by the power of $\zeta$ when
$m^2=0$ or Bessel functions when $m^2>0$.
The existence of the unstable mode corresponding the
modified Bessel function depends on the sign of
${\partial^2 W \over \partial\phi^2}$ at $z=z_{\rm vis}$.
Since
\be
\label{phP9b}
\left.{\partial^2 W \over \partial\phi^2}\right|_{z=z_{\rm vis}}
 \rightarrow - \sqrt{{\cal N} \over 2}g {a_2 \over 4}\e^{a_2\phi \over 2}
 <0\ ,
\ee
for large positive $z_{\rm vis}$, the sign is negative,
which is different from the case $z_{\rm vis} \rightarrow
-\infty$.
Therefore, remarkably there does not exist
an unstable mode corresponding to $m^2<0$ when we choose
$z_{\rm vis}\rightarrow +\infty$, where the cosmological
constant becomes very small. To conclude, the above analysis shows
the possibility to get the localized gravity near the brane corresponding
to our black hole solution. Note that excellent introduction to
universal aspects of gravity localization in brane world can be found in
ref.\cite{CEHS}.

\section{Newton law correction}

It is interesting to discuss now the
 correction to the Newton law coming from
KK mode.

First we consider $z_{\rm vis}\rightarrow -\infty$ case in
(\ref{Mp6}).
The constants of integration should
be determined by the boundary condition at $z=z_{\rm vis}$
in (\ref{MP7}).
As the overall factor $\sqrt{2\over \pi m \zeta}$ should be
absorbed into the measure for the normalization, we replace it
with unity and we impose a constraint as follows:
\be
\label{c2}
c_{1}^2+c_{2}^2 =1.
\ee
Then when $d=4$, and $m\zeta$ is large, $\psi(\zeta)$ behaves as
\be
\label{Ps}
\psi(\zeta)\sim c_{1}\mbox{cos}
\left(m\zeta -{(4\sqrt{3}+1)\pi\over 4} \right) +c_{2}\mbox{sin}
\left(m\zeta -{(4\sqrt{3}+1)\pi\over 4} \right)
\ee
so then $\partial_{\zeta}\psi(\zeta)$ is given by
\be
\partial_{\zeta}\psi(\zeta) \sim
-mc_{1}\mbox{sin}
\left(m\zeta -{(4\sqrt{3}+1)\pi\over 4} \right) +mc_{2}\mbox{cos}
\left(m\zeta -{(4\sqrt{3}+1)\pi\over 4} \right).
\ee
We should note that there might be some ambiguities in the
limiting procedure. Since we consider $\zeta_{\rm vis}
\rightarrow +\infty$, we first put $m\zeta$ to be large and
after that we choose the mass $m$ in the KK mode to be small,
which is relevant to the long range force.
Then by using the boundary condition (\ref{MP7}) and
the constraint (\ref{c2}),
we get $c_{1},c_{2}$ in following forms:
\bea
\label{c1c2}
c_{1} &=& \pm \kappa \sqrt{1 \over \kappa ^{2} +1} \nn
c_{2} &=& \pm \sqrt{1 \over \kappa ^{2} +1}
\eea
Here $\kappa $ is defined by
\bea
\label{kapa}
\kappa &\equiv & {-\mbox{cos}\zeta' + Z(\zeta =\zeta_{\rm vis})
\mbox{sin}\zeta' \over -\mbox{sin}\zeta' - Z(\zeta =\zeta_{\rm vis})
\mbox{cos}\zeta' } \nn
Z(\zeta =\zeta_{\rm vis}) &\equiv & {2\sqrt{2}
e^{A(\zeta=\zeta_{\rm vis})}W(\phi(z=z_{\rm vis}))\over m} \nn
\zeta' &\equiv &  m\zeta_{\rm vis} -{(4\sqrt{3}+1)\pi \over 4}
\eea
One can find the value of $\psi $ (\ref{Ps}) at
$\zeta=\zeta_{\rm vis}$  using (\ref{c1c2}), (\ref{kapa}):
\bea
\psi(\zeta_{\rm vis})&\sim &
\sqrt{1 \over \kappa ^{2} +1}
\left(\kappa ~\mbox{cos}\zeta'+\mbox{sin}\zeta'\right) \nn
&=& \sqrt{1 \over \kappa ^{2} +1}
\left( {1 \over \mbox{sin}\zeta' +Z(\zeta =\zeta_{\rm vis})
\mbox{cos}\zeta' } \right) \nn
&=& {\mbox{sin}\zeta'
+Z(\zeta =\zeta_{\rm vis})
\mbox{cos}\zeta'  \over \sqrt{1+Z(\zeta =\zeta_{\rm vis})^{2}} }
\left( {1 \over \mbox{sin}\zeta' +Z(\zeta =\zeta_{\rm vis})
\mbox{cos}\zeta' } \right) \nn
&=& {1 \over \sqrt{1+Z(\zeta =\zeta_{\rm vis})^{2}} }
\eea
If we take $m$ is small (or $|Z|$ is large), then
$\psi(\zeta_{\rm vis})$ is given by
\bea
\psi \sim {1 \over Z(\zeta =\zeta_{\rm vis}) }
={m \over 2\sqrt{2}
\e^{A(\zeta=\zeta_{\rm vis})}W(\phi(z=z_{\rm vis}))}
\eea
Then the correction to Newton's Law is
\bea
V(r) &\sim & \tilde G_4 {m_{1}m_{2}\over r}+\int_{0}^{\infty} dm
G_5{m_{1}m_{2}e^{-mr}\over r}\psi(\zeta =\zeta_{\rm vis})^2 \nn
&=&  \tilde G_4 {m_{1}m_{2}\over r}+\int_{0}^{\infty} dm
G_5 {m_{1}m_{2}e^{-mr}\over r}{m^2 \over 8
\e^{2A(\zeta=\zeta_{\rm vis})}W(\phi(z=z_{\rm vis}))^{2}}  \nn
&=& \tilde G_4{m_{1}m_{2}\over r}\left(1+
{G_5 \over \tilde G_4} {1 \over r^{3} 4
\e^{2A(\zeta=\zeta_{\rm vis})}W(\phi(z=z_{\rm vis}))^{2}}\right)
\eea
We should note that that the correction is not given by
${1 \over r^3}$ as in \cite{RS} but ${1 \over r^4}$.
This is mainly due to the limiting procedure where we first
have put $m\zeta$ to be large and after that we have chosen
the mass $m$ in the KK mode to be small.

Next we consider the case $z_{\rm vis}\to \infty$ given
in (\ref{PMxix}).
When $\zeta$ ($z$) is large, the Bessel functions
$J_\nu(-m\zeta)$ and $N_\nu(-m\zeta)$ behave as
\bea
\label{PMxxib}
J_\nu(-m\zeta)&\sim& \sqrt{-{2 \over \pi m\zeta}}
\cos\left(-m\zeta - {(2\nu +1)\pi \over 4} \right)\ ,\nn
N_\nu(-m\zeta)&\sim& \sqrt{-{2 \over \pi m\zeta}}
\sin\left(-m\zeta - {(2\nu +1)\pi \over 4} \right)\ .
\eea
For $d=D-1=4$ and ${\cal N}=1$ model in \cite{CLP},
we have the parameters $b$, $\nu$ given in (\ref{PMxxii}).
Then $c_1$ and $c_2$ in (\ref{PMxix})  should
be determined by the boundary condition at $z=z_{\rm vis}$ in
(\ref{MP7}) and constraint (\ref{c2}) again.
Since
\be
\partial_{\zeta}\psi(\zeta) \sim mc_{1}\mbox{sin}
\left(-m\zeta -{(2\nu +1)\pi \over 4} \right) -mc_{2}\mbox{cos}
\left(-m\zeta -{(2\nu +1)\pi \over 4} \right)\ ,
\ee
we can get $c_{1},c_{2} $ in the same way as (\ref{c1c2}).
\bea
\label{nc1c2}
c_{1} &=& \pm \kappa \sqrt{1 \over \kappa ^{2} +1} \nn
c_{2} &=& \pm \sqrt{1 \over \kappa ^{2} +1}
\eea
Here $\kappa $ is defined by
\bea
\label{kapa2}
\kappa &\equiv & {-\mbox{cos}\zeta' + Z(\zeta =\zeta_{\rm vis})
\mbox{sin}\zeta' \over -\mbox{sin}\zeta' - Z(\zeta =\zeta_{\rm vis})
\mbox{cos}\zeta' } \nn
Z(\zeta =\zeta_{\rm vis}) &\equiv & {2\sqrt{2}
e^{A(\zeta=\zeta_{\rm vis})}W(\phi(z=z_{\rm vis}))\over -m} \nn
\zeta' &\equiv &  -m\zeta -{(2\nu +1)\pi \over 4}
\eea
And $\psi$ (\ref{PMxix}) is written by using (\ref{nc1c2}),
(\ref{kapa2}).
\bea
\psi(\zeta)&\sim & \sqrt{1 \over \kappa ^{2} +1}
\left(\kappa ~\mbox{cos}\zeta'+\mbox{sin}\zeta'\right) \nn
&=& {1 \over \sqrt{1+Z(\zeta =\zeta_{\rm vis})^{2}} }
\eea
Then the correction to Newton's law in the limit that
$m$ is small (or $|Z|$ is large) is
\bea
V(r) &\sim & \tilde G_4 {m_{1}m_{2}\over r}+\int_{0}^{\infty} dm
G_5 {m_{1}m_{2}e^{-mr}\over r}\psi ^2 \nn
&=& \tilde G_{N}{m_{1}m_{2}\over r}+\int_{0}^{\infty} dm
G_5 {m_{1}m_{2}e^{-mr}\over r}{ m^2 \over 8
e^{2A(\zeta=\zeta_{\rm vis})}W(\phi(z=z_{\rm vis}))^{2}}  \nn
&=& \tilde G_4{m_{1}m_{2}\over r}\left(1
+{G_5 \over \tilde G_4}{1 \over 4
e^{2A(\zeta=\zeta_{\rm vis})}W(\phi(z=z_{\rm vis}))^{2}}
{1 \over r^3}\right)\ .
\eea
This is qualitatively the same type of correction as in ref.\cite{RS}.
Thus, the observer living on the brane Universe does not see drastic
changes in the Newton law.

\section{Dilatonic de Sitter brane Universe}

When $V$ is constant, instead of (\ref{Vi}), the solution
is given in \cite{NOtwo,NOcurv}, as follows
\bea
\label{curv2}
ds^2&=&f(y)dy^2 + y\sum_{i,j=0}^{d-1}\tilde g_{ij}(x^k)dx^i dx^j \nn
f&=&{d(d-1)  \over 4y^2
\lambda^2 \left(1 + { c^2 \over 2\lambda^2 y^d}
+ {kd \over \lambda^2 y}\right)} \nn
\phi&=&c\int dy \sqrt{{d(d-1) \over
4y^{d+2}\lambda^2 \left(1 + { c^2 \over 2\lambda^2 y^d}
+ {kd \over \lambda^2 y}\right)}}\ .
\eea
Here we define $V=-\lambda^2$ and
$g_{ij}$ is the metric of the Einstein manifold, which is
defined by $r_{ij}=k\tilde g_{ij}$, where $r_{ij}$ is the Ricci tensor
constructed with $\tilde g_{ij}$ and $k$ is a constant. Especially when $k$
is
given by $k=d\hat c^2>0$,
the metric on the brane can correspond to
the cosmological solution
\be
\label{c1}
\sum_{i,j=0}^{d-1}\tilde g_{ij}(x^k)dx^i dx^j
={1 \over \hat c^2t^2}\left(- dt^2 + \sum_{i=1}^{d-1}
\left(dx^i\right)^2\right)\ .
\ee
This solution describes the wall expanding and travelling in
5d universe. It may be presented as regular solution in terms of radial
coordinate similarly to black hole.

If one defines a new coordinate $z$ by
\be
\label{c2b}
z=\int dy\sqrt{d(d-1)  \over 4y^2
\lambda^2 \left(1 + { c^2 \over 2\lambda^2 y^d}
+ {kd \over \lambda^2 y}\right)}
\ee
and solves $y$ with respect to $z$, we obtain the warp
factor  $\e^{2A}=y(z)$.
We should note that there is a curvature singularity at
$y=0$ \cite{NOtwo,NOcurv}. Therefore we cannot put only one
brane but two branes and consider the region sandwiched by
them to avoid the singularity.

Using (\ref{Eivb}), one finds the
vacuum energy on the brane as follows
\bea
\label{c3}
U_{\rm hid}&=&-4\lambda \sqrt{{d-1 \over d}\left(
1 + { c^2 \over 2\lambda^2 y_{\rm hid}^d}
+ {kd \over \lambda^2 y_{\rm hid}}\right)} \nn
U_{\rm vis}&=&4\lambda \sqrt{{d-1 \over d}\left(
1 + { c^2 \over 2\lambda^2 y_{\rm vis}^d}
+ {kd \over \lambda^2 y_{\rm vis}}\right)} \ .
\eea
Here we assume $z_{\rm vis}>z_{\rm hid}$ and consider the
region $z_{\rm vis}\geq z\geq z_{\rm hid}$. $y_{\rm hid,vis}$
is the value of $y$ corresponding to $z=z_{\rm hid,vis}$.
 From (\ref{Eivb}), one gets
\be
\label{c4}
{\partial U_{\rm hid} \over \partial \phi}
= {2c  \over y_{\rm hid}^{d \over 2}}\ ,\quad
{\partial U_{\rm vis} \over \partial \phi}
= -{2c  \over y_{\rm vis}^{d \over 2}} \ ,
\ee
which tells that the vacuum energies $U_{\rm vis}$ on the brane
depend on the dilaton field.

One can consider the perturbation around the solution (\ref{curv2}),
(\ref{c1}). As it is difficult to consider the general case, we
first investigate the region $y<y_{\rm vis}$ and
$y_{\rm vis}\rightarrow 0$, that is, the brane is near
the singularity. We should note, however, it is sufficient to
consider the asymptotic region to check the existence
of the normalized zero mode and continuous KK modes.
Using (\ref{c2b}), when $y\rightarrow 0$, one gets
\be
\label{c5}
\zeta\sim {1 \over c}\sqrt{d \over d-1} y^{d-1 \over 2}\ .
\ee
This tells that we now consider the region where $\zeta$ is
small.
Since $A={1 \over 2}\ln y$, Eq.(\ref{PMivb}) has the
following form :
\be
\label{PMivbb}
0=m^2\psi + \partial_\zeta^2 \psi
+ {d-5 \over d-1}{1 \over \zeta} \partial_\zeta \psi
+ {4 \over (d-1)^2}{1 \over \zeta^2}\psi\ .
\ee
When $m^2=0$, the solution is given by
\be
\label{c0s}
\psi_0=\zeta^{2 \over d-1}
\left(\tilde c_1 + \tilde c_2 \ln\zeta\right)\ .
\ee
On the other hand, when $m^2>0$, we find
\be
\label{cms}
\psi_m=\zeta^{2 \over d-1}\left(c_1 J_0(m\zeta)
+ c_2 N_0(m\zeta)\right)\ .
\ee
The mode corresponding to $m=0$ does not seem to be normalizable
as in flat dilatonic brane \cite{GJS}. Then in order to localize the
gravity, we
need to put a brane corresponding to the hidden sector.

\section{Quantum effective action for dilatonic brane and RG flow
of Newton constant}

Our starting point is again the following action of 5d dilatonic
gravity (gauged supergravity):
\be
\label{Sa1}
S= {1 \over 16\pi}\left[ \int_M d^{d+1}x \sqrt{-g}\left( R
 - {1 \over 2}\partial_\mu\phi \partial^\mu\phi
 - V(\phi) \right) \right]\ .
\ee
The dilatonic potential is not specified for the moment.
As it was discussed in section 2,
the very common choice for $V$ is the exponential of dilaton:
It corresponds to the effective action for the breathing-mode
scalar and gravity which follows from KK sphere reduction from
M-theory or strings \cite{MSB}.
For the exponential potentials there are singular domain
wall solutions in above theory \cite{CLP}.
It is interesting that there are usually problems with
localization of 4d gravity when using only the action
(\ref{Sa1}).
As a result one should consider the inclusion of four
dimensional action (for walls they correspond to wall
source terms).
Then
\be
\label{Sa2}
S_{\rm source}=\sum_{i={\rm hid,vis}}\int d^4x \sqrt{-\gamma}
\left\{ L_{\rm QFT}\e^{\alpha_i\phi} + U_i(\phi)\right\}\ .
\ee
Here $U_i(\phi)$ are vacuum energies on branes, normally
$U_i(\phi)$ are dictated by the form of dilatonic potential.
For example, if $V\sim \e^{\kappa\phi}$ then $U_i$ has also
the exponential form. $L_{\rm QFT}$ is an arbitrary
Lagrangian corresponding to massless QFT (say, QED, QCD, SM,
GUT) which is classically conformally invariant in the background
$\tilde g_{\mu\nu}$. There is dilaton coupling on the brane which
is typical for Brans-Dicke gravity. Usually it is assumed to be
invisible in 4d world.

We are going to search for the solutions of the sort
\be
\label{Sa3}
ds^2 = dz^2 + \e^{2A(z)}\tilde g_{ij} dx^i dx^j
\ee
where 4 dimensional $\tilde g_{ij}=a^2(\eta)\eta_{ij}$.
It is assumed that branes sit on $z=z_{\rm hid}$ and
$z=z_{\rm vis}$.

We integrate over quantum fields in theory (\ref{Sa2}).
Supposing that there is only gravitational background,
the interaction of dilaton coupled quantum fields
leads to effective action induced by conformal anomaly
\cite{NNO}:
\be
\label{Sa4}
\Gamma_{\rm source}=\sum_{i={\rm hid,vis}} V_3\int d\eta
\left\{2b_1\sigma_1\sigma_1''''
 - 2(b_1+b)\left(\sigma_1'' - {\sigma_1'}^2\right)^2\right\}
\ee
where $\sigma=\ln a(\eta)$, $\sigma_1 = A + \sigma
+ {\alpha_1 \phi \over 3}$. For the sake of simplicity, we
adopt the large $N$-expansion (that justifies the neglecting
of proper quantum gravity contribution to (\ref{Sa4})).
If spinors give the leading contribution then
$b={3N_{1 \over 2} \over 60(4\pi)^2}$,
$b_1=-{11 N_{1 \over 2} \over 360 (4\pi)^2}$.
One can take the contribution above as
corresponding to maximally SUSY Yang-Mills theory (which
only changes the coefficients of above effective action).
That corresponds to implementing above compactification to AdS/CFT scheme.
Note that the suggestion to take into account the boundary matter
quantum effects (via conformal anomaly induced effective action for SUSY Yang-
Mills theory) in brane-world scenario has appeared in ref.\cite{NOZ}.
It was shown the possibility of creation of
de Sitter or Anti-de Sitter 4d Universe in 5d AdS space. In ref.
\cite{Hawking} the same idea  on application
of conformal anomaly has been expressed and effective brane tension
 due to such boundary quantum contribution for 4d de Sitter world has been
found.

Thus our complete action will be given by sum of three terms:
\be
\label{Sa5}
S_{\rm complete}=S + \Gamma_{\rm source}
+ \sum_{i={\rm hid,vis}}V_3\int d\eta \e^{4A}a^4(\eta) U_i(\phi)\ .
\ee

One can now consider the solution of the equations of motion given
from the action (\ref{Sa5}). In the bulk 5d universe, the action is
identical with the previous one (\ref{S}), then the solutions in
the bulk are also given by the previous ones. Especially when
$V(\phi)$ is a constant, we obtain the solution in (\ref{curv2}).
Near the brane, however, one obtains the following equations for
$d=D-1=4$
instead of (\ref{Eiv}) :
\bea
\label{Eivq}
\phi'' &\sim& \left[{\partial U_i(\phi)\over \partial\phi}
+ {\alpha_1 \over 3}\e^{4A}\left\{4b_1\sigma_1''''
- 4(b+b_1)(\sigma_1'''' - 6{\sigma_1'}^2\sigma_1'')\right\}
\right] \delta (z-z_i)\ , \nn
\quad A'' &\sim& -{1 \over 6}\left\{U_i(\phi)
4b_1\sigma_1''''
- 4(b+b_1)(\sigma_1'''' - 6{\sigma_1'}^2\sigma_1'')
\right\}\delta (z-z_i)\ .
\eea
Then by substituting the solution in (\ref{curv2}), (\ref{c2b}),
we find $U_i$ and ${\partial U_i \over \partial\phi}$ become
time dependent:
\bea
\label{c3b}
U_{\rm hid}&=&-4\lambda \sqrt{{3 \over 4}\left(
1 + { c^2 \over 2\lambda^2 y_{\rm hid}^4}
+ {4k \over \lambda^2 y_{\rm hid}}\right)}
 - {6b_1 y_{\rm hid}^2 \over t^4}\nn
U_{\rm vis}&=&4\lambda \sqrt{{3 \over 4}\left(
1 + { c^2 \over 2\lambda^2 y_{\rm vis}^4}
+ {4k \over \lambda^2 y_{\rm vis}}\right)}
 - {6b_1 y_{\rm vis}^2 \over t^4}\nn
{\partial U_{\rm hid} \over \partial \phi}
&=& {2c  \over y_{\rm hid}^{d \over 2}}
 - {8\alpha_1 b_1 y_{\rm hid}^2 \over t^4} \nn
{\partial U_{\rm vis} \over \partial \phi}
&=& -{2c  \over y_{\rm vis}^{d \over 2}}
 - {8\alpha_1 b_1 y_{\rm vis}^2 \over t^4}\ .
\eea

Hence, the price one paids for keeping the same de Sitter brane-world
solution is in change of brane vacuum energies. Quantum corrections
explicitly give contribution to vacuum energies which become
time-dependent (or dependent from the radius of de Sitter space in radial
coordinates). That indicates that effective brane tension will be changed.

Let us discuss now RG flow of 4d Newton constant
(for a recent review of holographic RG, see \cite{S}).
 Using (\ref{schw1}) for the solution in
(\ref{curv2}) for $d=4$, we find
\be
\label{schw1bb}
{1 \over \tilde G_4}
={1 \over G_5}\int_{y_{\rm hid}}^{y_{\rm vis}} dy {\sqrt{3}
\over \lambda \sqrt{1 + {c^2 \over 2\lambda^2 y^4}
+ {kd \over \lambda^2 y}}}\ .
\ee
If we define $U$ by $y_{\rm vis}=U^2$, we can identify
$U$ with the energy scale on the visible brane from AdS/CFT
correspondence \cite{W,Mal,GKP}. Therefore Eq.(\ref{schw1bb})
expresses the scale dependence of the gravitational coupling.
When $U$ ($y_{\rm vis}$) is small, we find
\be
\label{rg1}
{1 \over \tilde G_4}\sim {1 \over G_5}
{\left(y_{\rm vis}^3 - y_{\rm hid}^3\right) \over c\sqrt{3}}
={1 \over G_5}
{\left(U^6 - y_{\rm hid}^3\right) \over c\sqrt{3}}  \ .
\ee
Therefore the gravitational coupling $\tilde G_4$ becomes large
in the IR region, which would be due to the curvature
singularity at $y=0$.
On the other hand, when $U$ ($y_{\rm vis}$) is large, we find
\be
\label{rg2}
{1 \over \tilde G_4}\sim {1 \over G_5}
{\sqrt{3} \over \lambda }\left(y_{\rm vis}
 + Y(y_{\rm hid})\right)
= {1 \over G_5}
{\sqrt{3} \over \lambda }\left(U^2  + Y(y_{\rm hid})\right)  \ .
\ee
Here $Y$ depends on $y_{\rm hid}$ but does not on $y_{\rm vis}$.
Eq.(\ref{rg2}) seems to tell that the gravitational
coupling $\tilde G_4$ becomes small in the UV region.

We can also consider the scale dependence of the Newton
constant for the black hole type solution in Section \ref{BHsec}.
If the scale $U=\e^{A(z=z_{\rm vis})}$ is small (IR region),
$z_{\rm vis}$ becomes negative and large.
Then from (\ref{Mp1b}) and (\ref{schw1}), we find
\bea
\label{BHRG1}
U&\sim&\e^{\gamma z_{\rm vis}}\ ,\nn
{1 \over \tilde G_4}&=& {1 \over G_5}\int^{z_{\rm vis}} dz
\e^{2A} \sim {1 \over G_5}\int^{z_{\rm vis}} dz
\e^{2\gamma z} = {1 \over G_5} {\e^{2\gamma z_{\rm vis}}
\over 2\gamma}= {1 \over G_5} {U^2 \over 2\gamma}\ .
\eea
Here we assumed $z_{\rm vis} > z_{\rm hid} \rightarrow -\infty$
and put the constant of the integration to vanish. Therefore
$\tilde G_4$ becomes large in the IR region.
On the other hand, when the scale $U=\e^{A(z=z_{\rm vis})}$ is
large (UV region), $z_{\rm vis}$ becomes positive  and large.
By using (\ref{PMv}), we obtain
\bea
\label{BHRG2}
U&\sim&\alpha z_{\rm vis}^\beta\ ,\nn
{1 \over \tilde G_4} &\sim& {1 \over G_5}\int^{z_{\rm vis}} dz
\alpha^2 z^{2\beta}
= {1 \over G_5} \left({\alpha^2 z_{\rm vis}^{2\beta+1}
\over 2\beta + 1}+ \tilde Y(y_{\rm hid})\right) \nn
&=& {1 \over G_5}\left( {U^{2 + {1 \over \beta}} \over (2\beta +1)
\alpha^{1 \over \beta}}+ \tilde Y(y_{\rm hid}) \right) \ .
\eea
Here $\tilde Y$ depends on $y_{\rm hid}$ but does not on $y_{\rm vis}$.
For $d=D-1=4$ and ${\cal N}=1$ model in \cite{CLP},
we have $2 + {1 \over \beta}= {18 \over 5}$. Therefore
the gravitational coupling $\tilde G_4$ becomes small
in the UV region again.
We should note that the parameters specifying the black hole on the
brane do not enter in the above expressions  (\ref{BHRG1}) and
(\ref{BHRG2}), which is significantly different from the case that
there is a black hole in the bulk \cite{GJS2}.

\section{Discussion}

In summary, we presented the family of brane-world solutions of
5d dilatonic gravity. This family includes flat brane with effectively
small cosmological constant, (anti) de Sitter and Nariai spaces,
and brane-world dilatonic black holes. The study of gravity perturbations
around such black holes
shows that 4d gravity may be trapped. Corrections to Newton law near branes
are calculated. The proposal to take into account brane matter
quantum effects is made.( Actually, such proposal presented earlier in refs.
\cite{NOZ,Hawking} helps to formulate the problem in terms of AdS/CFT.)
The corresponding anomaly induced effective action is used to estimate
the role of quantum effects in realization of de Sitter branes in
asymptotically AdS dilatonic space. It is demonstrated that quantum
corrections change the brane vacuum energies. RG flow of Newton constant
in IR and UV is discussed.

There are many related problems which are left for future study. In
particular, it looks that the picture under discussion should be realized
within AdS/CFT correspondence. As it has been partially demonstrated
it is possible to do (at least in case of de Sitter brane). However,
the details of such quantum corrected brane-world Universe should be
investigated deeply.  The research of the role of brane matter
quantum effects to black holes is also extremely interesting topic.
Another open problem is the correct interpretation of dilatonic
brane-world black holes as holographic RG flows. The presented example of RG
flow of 4d Newton constant for dilatonic black hole represents the modest
 step in this direction.

\section{Acknoweledgements.}
We thank M.Ryan for the interest in this work. This research 
has been supported in part by CONACyT grant 2845E.

\end{document}